# On the Spatial Scaling of Seismicity Rate

G. Molchan [1,2] and T. Kronrod [1]

*1 International Institute of Earthquake Prediction Theory and Mathematical Geophysics, Russian Academy of Sciences, Warshavskoye sh. 79, kor.2 Moscow, 113556, Russia.*
*2 The Abdus Salam International Centre for Theoretical Physics, Trieste, Italy.*
E-mail: molchan@mitp.ru, kronrod@mitp.ru

Scaling analysis of seismicity in the space-time-magnitude domain very often starts from the relation $\lambda(m,L) = a_L 10^{-bm} L^c$ for the rate of seismic events of magnitude $M > m$ in an area of size $L$. There is some evidence in favor of multifractality being present in seismicity. In this case the optimal choice of the scale exponent $c$ is not unique. It is shown how different $c$'s are related to different types of spatial averaging applied to $\lambda(m,L)$ and what are the $c$'s for which the distributions of $a_L$ best agree for small $L$. Theoretical analysis is tested using California data.

## 1. Introduction

Certain classical laws of seismicity, namely, the Omori law for aftershocks and the Gutenberg-Richter (G-R) relation for earthquake energy, are reflections of one-dimensional properties exhibited by self-similar seismicity. Recent studies (see Bak et al., 2002; Corral, 2003, 2004; Baiesi and Paszuski, 2004b; Tosi et al., 2004; Abe and Suzuki, 2004) search for similar properties of earthquakes in the multidimensional phase space of location-time-magnitude. The results obtained along this line of research introduce substantial changes into the conventional notions of seismicity; therefore, analyrification are in order.

The pioneering work of Bak et al. (2002) is concerned with the distribution of interevent time $\tau_L$ between successive events of magnitude $M > m$ in a $L \times L$ cell. These authors averaged observed distributions of $\tau_L$ over all $L \times L$ cells of the grid covering a seismic zone $G$ to find that

$$P\{\tau_L < t\} = F_\tau(t\lambda_L) \qquad (1)$$

where $F_\tau$ is a unified (i.e., independent of $m$ and $L$) function, and $\lambda_L$ is a measure of $M > m$ events per unit time in an $L \times L$ cell. The contribution of a cell into the averaging result is proportional to the number of such events in the cell. The following representation is used:

$$\lambda_L = a 10^{-bm} L^c, \qquad (2)$$

where the quantity $b$ is the slope in the Gutenberg-Richter law, while the exponent $c$ is associated with a fractal dimension of earthquake epicenters, i.e., $0 \le c \le 2$.

Keilis-Borok et al. (1989) seem to have been the first to put forward relation (2) as a generalization of the G-R law for a "typical" area of size $L$, when seismic events are subject to fractal geometry. Viewed as such, (2) gives rise to several queries:

– Why is it that the parameter $c$ is independent of $m$?
– If $c$ is a fractal dimension, exactly which one is it?

There is a whole one-parameter family of the so-called Renyi dimensions or generalized dimensions in the terminology of Grassberger and Procaccia (1983), $d_q$, $q \ge 0$. The most popular of these are the correlation dimension $d_2$ and the capacity/box dimension $d_0$. The procedure of estimating the exponent leads to the correlation dimension both in the original work (Keilis-Borok et al., 1989) and in a later follow-up study (Kossobokov and Nekrasova, 2004), although the reasons for introducing (2) are based on the box dimension. Some workers are using (2) for the same purposes in one and the same area (California) with different exponents $c$: $d_2 = 1.2$ and $d_0 = 1.6$ (see Bak et al., 2002, and Corral, 2003, or else Baiesi and Paszuski: 2004a and 2004b).

The situation gets more complicated, when the reasoning of Pisarenko and Golubeva (1996) is considered. These workers start from the hypothesis of self-similarity for seismicity in space-time and develop a model where $M > m$ events form a Poisson process in any subregion $A$ of region $G$ with a random rate $\lambda(A)$. To be more specific, the set function $\lambda(A)$ is a sample of a random Levy measure, i.e., a measure with independent increments and a stable distribution of the index $0 < \alpha < 1$. The case $\alpha \ge 1$ is impossible, because $\lambda(A)$ is positive. In this model one has for every $L \times L$ cell:

$$\lambda(L \times L) = a_L 10^{-bm} L^c, \qquad (3)$$

where $c = 2/\alpha > 2$. (A formal averaging of $\lambda(L \times L)$ over the cells must lead to relation (2) with the average parameter $a$). Pisarenko and Golubeva (1996) give $\alpha = 0.57$ for southern California, from which one gets a most unusual value, $c = 3.5$. From the model it follows that the population of $\{a_L\}$ obeys a unified distribution, namely, the stable Levy distribution with index $\alpha < 1$. Consequently, $P(a_L > x) \sim cx^{-\alpha}$ for $x \gg 1$. In that case, however, $a_L$ does not have the ordinary mean value, and so the parameter $a$ in (2) may become meaningless within the framework of the model considered.

The above model is of interest in that it suggests a generalized frequency-magnitude relation of a more flexible form than (2). It is specified by (3) with a unified distribution of normalized quantities $\lambda(L \times L)/\lambda_L$, i.e., one has the exact equality

$$P\{\lambda(L \times L) < x\} = F_\lambda(x/\lambda_L), \qquad (4)$$

with the unified function $F_\lambda$ involved. That circumstance is not trivial, since the analogous relation (1) can hold approximately only. If (1) holds exactly, the spatial rate of $M > m$ events must be constant in $G$ (see Molchan, 2005).

In an independent study, Corral (2003) derived (4) for California at a fixed exponent $c = 1.6$. He got $F_\lambda(x) = 0(x^{-1.2}), x \gg 1$, which differs from the above model in which $F_\lambda(x) = 0(x^{-0.57})$ and $c = 3.5$.

It thus appears that the choice of a suitable exponent $c$ for deriving unified seismicity laws remains an open question. The phrase *suitable exponent* will be interpreted as follows: find a value of $c$ such as to make observed distributions of normalized scale-dependent quantities ($\tau_L \lambda_L$ or $\lambda(L \times L)/\lambda_L$, say) sufficiently close to one another for small $L$. Considering (2) as a G-R law for *typical* area of size $L$, then the notion of a suitable for $c$ and typical for a cell must be made consistent among themselves.

Below the choice of $c$ will be discussed for the statistic $\lambda(L\times L)$. In that case theory can predict certain things, if one assumes multifractality for the measure $\lambda(dg|m)$ which gives the rate of $M > m$ events per unit time in an area $dg$. For measures of this type the generalized dimensions $d_q$ are defined, and some of them are not identical. It should be admitted at once that the hypothesis of multifractality for seismicity has both adherents (Geilikman et al., 1990; Hirata and Imoto, 1991; Hirobayashi et al., 1992; Godano et al., 1999) and opponents (Gonzato et al., 1998; Eneva, 1996). This is not surprising. Multifractality is a sophisticated idealization of physical objects demonstrating a diversity of local similarity, the full spectrum of a multifractal is difficult to measure; besides, multifractal objects are not always easily distinguishable from pseudo-fractals or their intermediate forms, even when extensive data is available (see examples in Gorski, 2001; Molchan and Turcotte, 2002). In practical terms, the best that can be done is to observe a multifractal behavior of the measure in a range of scales $\Delta L$. In that case the quantity $c$ in (2) becomes a parameter for the unified law under study in the range $\Delta L$. Below, our theoretical analysis of the populations $\{\lambda(L\times L)\}$ will be supplemented with an analysis of California seismicity.

## 2. Scalings for multifractal seismicity.
### 2.1. The measure $\lambda(dg|m)$ as a multifractal.

Apply a rectangular grid of step $L$ to region $G$ to partition it into $L\times L$ cells. Denote the rate of $M > m$ events in $G$ as $\lambda(G)$, and that in the $L\times L$ cell as $\lambda(L\times L)$. The number of cells with $\lambda(L\times L)$ positive is denoted $n(L)$. If the following relation holds:

$$\lg n(L) = -d_0 \lg L\,(1+o(1)),\ \ L\to 0,\ 0<d<2 \qquad (5)$$

then the support of $\lambda(dg|m)$ is said to be fractal and to have a box dimension $d_0$. When $\lambda(dg|m)$ is multifractal, the support stratifies as it were into a sum of fractal subsets $S_\alpha$ having the Hausdorff dimensions $f(\alpha)$. Taking any point in $S_\alpha$, there exists a sequence of areas $L\times L$, $L\to 0$ such that

$$\lg \lambda(L\times L) = \alpha \lg L\,(1+o(1)). \qquad (6)$$

Relation (6) describes a type of spatial concentration of events or a type of singularity for $\lambda(dg|m)$; the parameter $\alpha$ itself is termed Hoelder's exponent or a local dimension of the measure. Accordingly, $f(\alpha)$ describes the Hausdorff dimension of points having the singularity type $\alpha$. The pairs $(\alpha, f(\alpha))$ form the multifractal spectrum of $\lambda(dg|m)$. The information on the multifractal behavior of $\lambda(dg|m)$ is derived from the Renyi function:

$$R_L(q) = \sum \lambda(L\times L)/\lambda(G))^q,\ \ |q|<\infty, \qquad (7)$$

Here and below, the summation is over all $L\times L$ squares with $\lambda(L\times L) > 0$. When multifractals are considered, the function is asymptotically

$$\lg R_L(q) = \tau(q)\lg L\cdot(1+o(1)),\ \ L\to 0, \qquad (8)$$

where the scaling index $\tau(q)$ is related to $f(\alpha)$ through the Legendre transformation:

$$\tau(q) = \min_\alpha(q\alpha - f(\alpha)). \qquad (9)$$

When $q = 0$, relation (8) becomes (5), hence $\tau(0) = -d_0$. In the case of a monofractal, where the spectrum $(\alpha, f(\alpha))$ degenerates to the point $(d_0, d_0)$, the function $\tau(q)$ is linear, $\tau(q) = d_0(q-1)$. In the general case $-\tau(q)$ is a convex function with $\tau(1) = 0$ (see the example in Fig. 2). If $-\tau(q)$ is strictly convex and smooth, the region of values of $\dot\tau(q)$ defines the interval of possible singularities of $\alpha$ in (6), while the Legendre transformation of $\tau(q)$:

$$f(\alpha) = \min_q(q\alpha - \tau(q))$$

describes the Hausdorff dimensions of these singularities. These statements constitute the content of multifractal formalism (see, e.g., Feder, 1988), which was found to hold for many mathematical examples. There are pathologic cases in which the function $-\tau(q)$ exists, but not is convex. Practically important examples of such pseudomultifractals can be found in Gorski (2001).

The quantities $d_q = \tau(q)/(q-1)$ are known as Renyi dimensions or as the generalized Grassberger-Procaccia dimensions. Because $(-\tau(q))$ is convex, the numbers $d_q$ do not increase with increasing $q$. From the relation $\tau(1) = 0$ and the mean value theorem one has

$$d_q = \frac{\tau(q)-\tau(1)}{q-1} = \dot\tau(q^*), \qquad (10)$$

where $q^*$ is a point between 1 and $q$. For this reason $d_q$ (in the case of smooth and strictly convex $-\tau(q)$) describes a type of singularities or a local dimension of $\lambda(dg|m)$. One has $d_1 = \dot\tau(1)$ when $q = 1$. That quantity is known as the information dimension, being remarkable because it is the root of $\alpha = f(\alpha)$. Corresponding to solutions of that equation are usually sets $S_\alpha$ of a positive $\lambda$–measure, hence these are the most interesting from the physics point of view. The strict convexity and smoothness of $-\tau(q)$ in a vicinity of 1 ensures that $\alpha = f(\alpha)$ has a single root. In that case the closure of the set $S_\alpha$, $\alpha = d_1$ defines the topologic support of $\lambda(dg|m)$. We shall assume a regular situation to be the case when the Hausdorff and box dimensions of the support are identical.

To sum up, it is only $d_0$ and $d_1$ of all generalized dimensions which are related to the fractality of a measure support, the others $d_p \neq d_0, d_1$ providing information on local types of measure singularity. Examples of theoretical analyses of multifractals, both deterministic and stochastic ones, can be found, e.g., in (Pesin and Weiss, 1997; Mandelbrot, 1989; Molchan, 1996).

### 2.2. Scaling the averages of $\lambda(L\times L)$.

In order to characterize the rate of $M > m$ events in region $G$ in an $L\times L$ cell, we average the $\lambda(L\times L)$ over all cells where $\lambda(L\times L) > 0$ using some weights. The choice of weights is governed by the goals for which we are going to use the average. The following one-parameter family of weights is sufficiently flexible and natural:

$$m^{(p)}(L\times L) = k_p \lambda^{(p)}(L\times L),\ \ |p|<\infty,\ \lambda(L\times L)>0,$$

where $k_p$ is a normalizing constant such that $\sum m^{(p)}(L\times L) = 1$, i.e., recalling (7),

$$1/k_p = R_L(p)\,\lambda^p(G).$$

When $p = 0$, we have the ordinary averaging of the $\lambda(L\times L)$ for $\lambda(L\times L) > 0$, while when $p\gg 1$, the average will characterize the most active cells, because

$$\sum \lambda(L\times L)m^{(p)}(L\times L) \to \max \lambda(L\times L),\ p\to\infty.$$

Denote the average with the weights $<.>_p$ as $m^{(p)}(L \times L)$. Then

$$<\lambda(L \times L)>_p = \sum \lambda(L \times L) m^{(p)}(L \times L) =$$
$$= \lambda(G) R_L(p+1)/R_L(p).$$

If (8) holds, then
$$\lg <\lambda(L \times L)>_p = [\tau(p+1) - \tau(p)] \lg L (1 + o(1)) + \lg \lambda(G)$$
or
$$<\lambda(L \times L)>_p \sim \lambda(G) L^{c_p}, \quad (11)$$

where $c_p$ has the nontrivial form
$$c_p = \tau(p+1) - \tau(p) = p d_{p+1} - (p-1) d_p. \quad (12)$$

The rate $\lambda(G)$ in a large region can be fairly well described by the Gutenberg-Richter magnitude-frequency relation $\lambda(G) = a 10^{-bm}$; consequently, (11, 12) constitute an updated variant of (2) for the case of a multifractal measure $\lambda(dg \mid m)$.

The averaging cases of most interest are $p = 0$ and $1$. Then
$$c_p = \begin{cases} d_0 & \text{box dimension}, \quad p = 0 \\ d_2 & \text{correlation dimension}, \quad p = 1. \end{cases}$$

Ordinary averaging $<\lambda(L \times L)>_0$ thus corresponds to the scaling index $c = d_0$, i.e., to the box dimension of the support of $\lambda(dg \mid m)$, while the averaging that is proportional to the rate of events in $L \times L$ corresponds to the correlation dimension $c = d_2$.

The weights $\{m^{(p)}(L \times L)\}$ can be interpreted as the probability distribution $P_L^{(p)}$ governing the sampling of $L \times L$ cells. In that case (11) describes the rate of $M > m$ events in a $P_L^{(p)}$ – random $L \times L$ cell in $G$. Similarly to (10), we conclude that
$$c_p = \tau(p+1) - \tau(p) = \dot{\tau}(p + \theta^*), \quad 0 \leq \theta^* \leq 1,$$
i.e., $c_p$ can correspond to some type of singularity for $\lambda(dg \mid m)$.

**2.3. Scaling the distribution of $\lambda(L \times L)$.**

Consider the population of normalized quantities $\lambda(L \times L)/\lambda_L$, i.e.,
$$\xi_L = \lambda(L \times L)/[\lambda(G)(L/L_0)^c], \quad (13)$$

related to the partitioning of region $G$ into $L \times L$ cells. Here $L_0$ is the external scale of region $G$, say, $L_0 = \sqrt{area\ of\ G}$ and the $\xi_L$ are different from $a_L$ in (3) by a constant factor. Corral (2003) found that the distribution of $\xi_L$ for California with $c = 1.6$ is practically independent of $L$ in the range 10-120 km and $m = 2$–3. Corral (2003) also asserts that the distribution of $\xi_L$ is weakly dependent on the choice of the time interval $\Delta T$ in the range of 1 day to 9 years. The assertion about $\Delta T$ needs to be made more specific in order to be reproducible. Nevertheless, one may pose the following question for multifractal measures $\lambda(dg \mid m)$: for what values of $c$ the distribution of $\xi_L$ has a limit as $L \to 0$? Similarly to Section 2.2, we will extend the problem using the weights $m^{(p)}(L \times L) = k_p \lambda^p(L \times L)$ as a probability measure $P_L^{(p)}$ for $\xi_L$. Taking the case $p = 0$, we then arrive at the distribution of $\xi_L$ treated by Corral (2003).

The class of multifractal measures is very broad, and the measures themselves may have rather complicated structure. For this reason we quote standard heuristic arguments in order to find a suitable $c = c^{(p)}$ for any $p$, so as to be able to expect a nontrivial limiting distribution for $(\xi_L, P_L^{(p)})$.

Denote the multifractal spectrum of $\lambda(dg \mid m)$ by $f(\alpha)$. Then the number of $L \times L$ cells of type $\alpha$, i.e., such that $\lambda(L \times L) \sim L^\alpha$, is increasing like $L^{-f(\alpha)}$. For this reason $\lambda(L \times L)/L^c$ is bounded away from 0 and $\infty$ as $L \to 0$, if the $L \times L$ cell belongs to the type $\alpha = c$. The probability or weight of a cell of type $\alpha$ is of order
$$L^{-f(\alpha)} m^{(p)}(L \times L) = L^{-f(\alpha)} \lambda^p(L \times L)/R_L(p)$$
$$\sim L^{-f(\alpha) + p\alpha}/L^{\tau(p)},$$

where $R_L(p)$ is given by (7), and $\tau(p)$ is by (9) equal to $\tau(p) = \min_\alpha(p\alpha - f(\alpha))$. The resulting probability is bounded away from 0, when $L \neq 0$, provided $\tau(p) = p\alpha - f(\alpha)$. It follows that the desired $c = c^{(p)}$ is such that the function $p\alpha - f(\alpha)$ reaches its minimum at $\alpha = c$; in short, $c^{(p)} = \arg \min_\alpha (p\alpha - f(\alpha))$.

In particular, when $p = 0$, the desired $c$ is the point of maximum for $f(\alpha)$, while when $p = 1$, it is identical with the information dimension $d_1$ for which $d_1 = f(d_1)$. Consequently, if $f(\alpha) = d_0$ and $f(\alpha) = \alpha$ have unique solutions, then
$$c^{(p)} = \begin{cases} \alpha : f(\alpha) = d_0, & p = 0 \\ d_1, & p = 1. \end{cases}$$

If the spectrum $f(\alpha)$ is a strictly convex function, it can be described parametrically in terms of $\tau(q)$: $\alpha = \dot{\tau}(q)$, $f(\alpha) = q\alpha - \tau(q)$. Hence (since $-\tau(q)$ is convex)
$$\alpha p - f(\alpha) = (p - q)\dot{\tau}(q) + \tau(q) \geq \tau(p).$$

The left-hand side reaches the minimum at $q = p$. Consequently,
$$c^{(p)} = \dot{\tau}(p). \quad (14)$$

Consider some examples. Let the measure $\lambda(dg \mid m)$ have the density $\dot{\lambda}(g \mid m)$; the spectrum $f(\alpha)$ then consists of the single point $(\alpha, f(\alpha)) = (2, 2)$, so that $c^{(p)} = d_0 = 2$. Indeed, we can make the following statement: the distributions $(\xi_L, P_L^{(p)})$ have limits as $L \to 0$. Namely, when $p = 0$,
$$\lim_{L \to 0} \frac{\#\{0 < \xi_L < x\}}{\#\{\xi_L > 0\}} =$$
$$\frac{mes\{g : 0 < L_0^2 \dot{\lambda}(g \mid m)/\lambda(G) < x\}}{mes\{g : \dot{\lambda}(g \mid m) > 0\}} = F_\lambda^{(0)}(x). \quad (15)$$

Here we use the notation $mes\{g : \varphi(g) < x\}$ for the area of points $\{g\}$ for which $\varphi(g) < x$. The limit is independent of the choice of the partition of $G$. When $p > 0$, the limit of $(\xi_L, P_L^{(p)})$ is
$$F_\lambda^{(p)}(x) = \int_0^x u^p dF_\lambda^{(0)}(u) / \int_0^\infty u^p dF_\lambda^{(0)}(u). \quad (16)$$

We now take up a more complicated example. Consider a measure $\lambda(dg \mid m)$ that has densities in the square $[0, 1]^2$ and on the interval $[1, 2]$. This is a bifractal mixture with two points in the spectrum $(\alpha, f(\alpha))$: $(2, 2)$ and $(1, 1)$. For this we have $\tau(q) = \min(d_0(q - 1), d_2(q - 1))$ where $d_0 = 2$ and $d_2 = 1$. We get $c^{(p)} = \dot{\tau}(p)$ i.e. $d_0$ when $0 \leq p < 1$, and $d_2$ when $p > 1$. In both of these cases there exist limiting measures that can be written down analogously to (15) and (16). They depend on the component of $\lambda(dg \mid m)$ in $[0, 1]^2$ when $c = 2$ and on that in $[1, 2]$ when $c = 1$. When $p = 1$, the situation is similar to phase transitions in thermodynamics. Two normalizations are possible, with $c = d_2$ and with $c = d_0$. When $c = d_2$ ($c = d_0$), the limiting distribution has both a density that is determined by $\lambda(dg \mid m)$ on $[1, 2]$ ($[0, 1]^2$) and the $\delta$–function concentrated at 0 ($\infty$), respectively.

Relations (12) and (14) point to an interesting fact, namely, the exponents $c$ that are suitable for scaling $\{\lambda(L \times L), P_L^{(p)}\}$ and its mean are generally not identical. We are going to show that

$$c_p \leq c^{(p)} \qquad (17)$$

This can be seen as follows. The function $-\tau(q)$ is a convex one, hence it lies above the chord that connects the points $(p, \tau(p))$ and $(p+1, \tau(p+1))$ in the interval $(p, p+1)$, i.e., $\tau(q) \geq \tau(p) + (\tau(p+1) - \tau(p))(q-p)$, $p < q < p+1$ In that case, however,

$$c^{(p)} = \lim_{q \to p} \frac{\tau(q) - \tau(p)}{q - p} \geq \tau(p+1) - \tau(p) = c_p.$$

Because $-\tau(q)$ is a convex function, the relation $c^{(p)} = c_p$ is possible, if $\tau(q)$ in $(p, p+1)$ is a linear function.

**2.4. Estimation of $\tau(q)$.**

The test area for the analysis of unified seismicity laws is the California catalog of $M \geq 2$ events. For this region we know the estimates $d_0 = 1.6$ (Corral, 2003) and $d_2 = 1.1 - 1.2$ (Kagan, 1991; Kossobokov and Nekrasova, 2004) which favor a nonlinear $\tau(q)$, $\tau(q) \neq d_0(q-1)$, hence indicate that the choice of $c$ in (2) is not unique. The same fact is corroborated by strictly decreasing dimensions $d_q = \tau(q)/(q-1)$ in the interval $2 \leq q \leq 5$ as found by Godano et al. (1999) from the $M \geq 1.5$ seismicity for the period 1975-1995. The numerical value $d_2 = 0.85$ in the last paper referred to is widely divergent from $d_2 = 1.2$ for $M \geq 2$. At the same time the above publications do not contain any information required for comparing the estimates of $d_q$. For this reason the nonlinearity of $\tau(q)$ calls for independent verification.

In formal terms $\tau(q)$ is defined through the Renyi function $R_L(q)$ (see (17)) as the slope of $\lg R_L(q)$ plotted against $\lg L$ for small $L$. The chief difficulty consists in finding the range of scale $(L_*, L^*)$ where the slope estimate is stable.

The necessity of the lower threshold $L_*$ is due to the fact that the set of seismic events is finite. The number of cells $n_L$ covering the support of $\lambda(dg \mid m)$ is of order $(L_0/L)^{d_0}$. If $L$ varies as a geometric progression, then $n_L$ is rapidly increasing as $L \to 0$. For this reason any nonempty $L \times L$ cell will contain 1 event beginning from some small $L$. This leads to the formally correct (but erroneous) estimate $d_q = 0$, $q \geq 0$. Goltz (1997) demands that the mean number of events per cell be $k$ ($k = 5$) or greater, i.e., $N/n_L \geq k$ where $N$ is the total number of events. In that case one arrives at the restriction $L_* \geq (k/N)^{1/d_0} L_0$. This will hold for all $d_0 \leq 2$, provided

$$L_* = \sqrt{k/N}\, L_0. \qquad (18)$$

The threshold (18) with $k = 1$ was proposed by Nerenberg and Essex (1990). It is rather coarse, being adapted to any measure $\lambda(dg \mid m)$. To overcome that drawback we note the following. Theoretically, isolated points in the support of $\lambda(dg \mid m)$ do not affect its multifractal spectrum or the generalized dimensions $d_q > 0$. For this reason we will also compute $R_L(q \mid k)$ along with the Renyi function $R_L(q)$. The empirical analogue of $R_L(q)$ sums up the nonzero values $[n(L \times L, T)/n(G, T)]^q$ over $L \times L$ cells; here $n(A, T)$ is the number of events in $A$ for the period $T$. Accordingly, $R_L(q \mid k)$ takes account of only those cells where $n(L \times L, T) > k$. The scale $L$ starting from which the functions $R_L(q)$ and $R_L(q \mid 1)$ begin to diverge substantially can naturally be taken as $L_*$. In other words, that scale is taken as $L_*$ below which the discrete (on the scale $L_*$) component of the support plays a significant part in estimation of the multifractal spectrum. The estimate of $L_*$ proposed above is efficient for small $q$, because the contribution of terms like $[1/n(G, T)]^q$ in $R_L(q)$ is rapidly decreasing with $q$ increasing ($q > 1$). The formal rule for choosing $L_*$, when $q \geq 0$, can be expressed in the form

$$\frac{|R_{L_*}(0) - R_{L_*}(0 \mid 1)|}{R_{L_*}(0)} < \varepsilon,$$

where $\varepsilon$ is a small parameter (below $\varepsilon = 10\%$).

Consider the upper bound $L^*$. The conventional estimation of $\tau(q)$ is based, not on the Renyi function, but on an integral modification of it:

$$I_L(q) = \frac{1}{N} \sum_{i=1}^{N} [n(B_L(g_i), T)/n(G, T)]^{q-1}, \; q \neq 0, 1. \qquad (19)$$

Here, $B_L(g)$ is a circle of radius $L$ centered at $g$, and $\{g_i\}$ are the epicenters of events in the catalog. To keep most of the circles $B_L(g_i)$ within region $G$, Nerenberg and Essex (1990) suggested using $L^* = \rho L_0$; one has $\rho = 1/6$, if region $G$ is nearly a circle and the measure $\lambda(dg \mid m)$ is nearly uniform. It thus appears that the motivation of the estimate of $L^*$ is related to the choice of the tool for estimating $\tau(q)$ rather than to the nature of the problem. The treatment of illustrative examples based on self-similar objects available in the literature uses $L^* = L_0$. However, the simplest possible fractal object (Cantor's staircase) disintegrates into similar parts only when one has a special choice of the scale, $L = (1/3)^k$, and of the observation interval. In practice therefore, it is natural to deal with scales for which the set $\{n(L \times L, T)\}$ with nonzero $n(L \times L, T)$ is not small, i.e., when the above set can be treated as a statistical population. If $n_L \sim (L_0/L)^{d_0}$, $n_L > 100$, and $d_0 \cong 2$, then $L^* \cong L_0/10$.

The estimate of $L^*$ proposed is also effective for small $p$. As p increases, the main contribution into $R_L(q)$ is due to points of high concentration; these are few in a limited data set, hence it is more difficult to make a representative statistical selection of $n(L \times L, T)$. This constitutes the chief obstacle for reliable estimation of $\tau(q)$ at large $q$. In turbulence (see Frisch, 1996) which supplies probably the best data for multifractal analysis, $\tau(q)$ was found to be nonlinear for $0 \leq q \leq 5$ in the energy dissipation field at large Reynolds numbers.

If $(\underline{L}, \overline{L})$ is a straight segment in the plots of $(\lg R_L(q), \lg L)$, $0 \leq q \leq q_0$, then we shall say that seismicity exhibits multifractal behavior in the range of scales $(\underline{L}, \overline{L})$. Gonzato et al. (1998) demand $\overline{L}/\underline{L} \geq 10^s$, $s \geq 3$ to make the above statement convincing. This is a stringent requirement from the standpoint of applications; in the case under consideration, it implies both multifractality and self-similarity of seismicity in a wide range of scales. If $L_0 = 10^3$ km, then $L^* \sim 100$ km, while when the epicenter location accuracy is $\Delta = 1$ km $\leq L_*$, then $\overline{L}/\underline{L} \leq 100$. The scale range $\Delta L = (10-120$ km$)$ is encountered in Corral's (2000) analysis of unified seismicity laws. It follows that the statement asserting a multifractal behavior of seismicity in this range $\Delta L$ is of interest for applications.

### 3. California seismicity.

*The data.* Following Corral (2003), we are going to examine observed 2D seismicity for the rectangular subarea in California: $G = (30°\text{N}, 40°\text{N}) \times (113°\text{W}, 123°\text{W})$ (see Fig. 1). The seismicity we use includes $m \geq 2$ events with depth of focus down to 100 km for the period 1984-2003. The ANSS catalog (2004) that we use contains 116,700 such event.

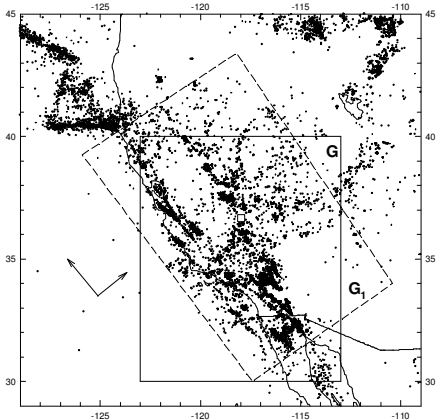

Figure 1. California: seismic events with $M \geq 3$ (●); center (◊) and principal directions (↑, →) of the rectangular grid; the main ($G$) and the alternative ($G_1$) seismic regions for dimension computations.

*The tau function* (Fig. 2). Stable values of $\tau(q)$ were obtained for $0 \leq q \leq 3$ and for the range of scales $\Delta L = (10-20, 100)$ km. The box counting method we used to estimate $\tau(q)$ is described in Section 2.4 and illustrated in Figs. 3 and 4(a, b). The axis of the grid which covers $G$ was made to lie along the San Andreas fault. An $L \times L$ cell was incorporated in the computation, if its center and at least three corners belonged to $G$. The estimation results are explained below.

The linear size of $G$ is $L_0 = \sqrt{area\ of\ G} = 1004$ km, hence $L^* = L_0/10 = 100$ km is a prior estimate for the upper bound of the scale. This choice is corroborated by the numbers of nonempty $L \times L$ cells. One has

| $L$ km | 10 | 20 | 40 | 80 | 100 |
|---|---|---|---|---|---|
| $n_L$ | 4366 | 1544 | 468 | 128 | 80 |

When $L = 160$ km, one has $n_L = 29$, which is not enough to consider $\{\lambda(L \times L)\}$ as a statistical population.

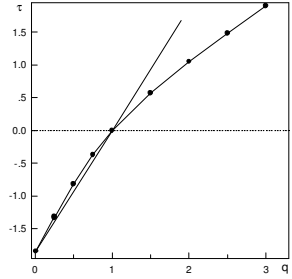

Figure 2. Tau-function for $M \geq 2$ events in region $G$ (see Fig. 1); it is based on the interval of scales $\Delta L = (10, 100)$ km. The straight line is $\tau(q)$ for a monofractal with the observed dimension $d_0 = -\tau(0)$. The numerical values of $\tau$ are shown in Figs. 3 and 4.

The box dimension $d_0 = -\tau(0)$ presents the greatest difficulty for estimation. The statistic $R_L(0|k)$ in Fig. 3 determines the number of $L \times L$ cells with numbers of events greater than $k$, $k = 0, 1, 2, 3,$ and 4. The curves of $\lg R_L(0|0)$ and $\lg R_L(0|1)$ significantly diverge when $L < 10$ km. This accounts for the choice of the lower bound $L_* = 10-20$ km for the $m \geq 2$ events. The slope in the plot of $(\lg R_L(0|0), \lg L)$ in the interval $(L_*, L^*)$ was estimated by least squares; the estimates of the box dimension are as follows: $d_0 = 1.82$, if $\Delta L = 20-100$ and 1.74, if $\Delta L = 10-100$ km.

Figure 3(b) contains similar data for estimating $d_0$ based on the $m \geq 3$ events. The interval $(L_*, L^*) = (20-40, 100)$ km is too narrow there for reliable estimation of $d_0$.

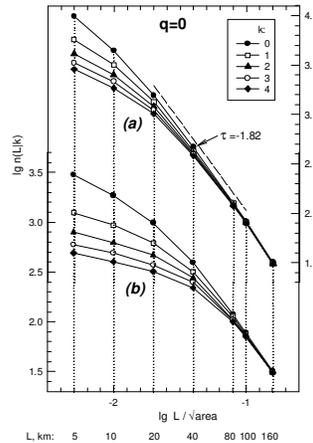

Figure 3. Data for estimating the box dimension $d_0 = -\tau(0)$ based on $M \geq 2$ events (a) and $M \geq 3$ events (b). The vertical axis shows the number of $L \times L$ cells, $n(L|k)$, that have numbers of events greater than $k$, $k = 0, 1, 2, 3, 4$. The dashed line shows both the slope $\tau(0)$ and the interval of scales $\Delta L$ for estimation of $\tau(0)$ by least squares using $n(L|0)$.

The estimation of $\tau(q)$ $q > 0$ calls for no additional explanation. Figure 4 (a, b) shows that the estimation of $\tau(q)$ is stable for $q = 0.25-2.5$ in the interval of scales $\Delta L = 10-100$ km. The stability is disturbed from the value $q = 3$ upwards (see Fig. 4b). The causes of this were discussed in Section 2.4. Computation of $\tau(q)$ at negative $q$ require high accuracy in the estimates of $\lambda(L \times L)$ in cells with low numbers of events. The requirement is not realistic for statistical reasons. Consequently, our estimates of $\tau(q)$ are for the interval $0 \leq q \leq 3$. The fact that $-\tau(q)$ is a convex function (Fig. 2) and that the plots in Figs. 3 and 4 are linear in the interval $\Delta L = (10, 100)$ km, entitles one to say that the measure $\lambda(dg | m = 2)$ looks like a multifractal in the above range of scales.

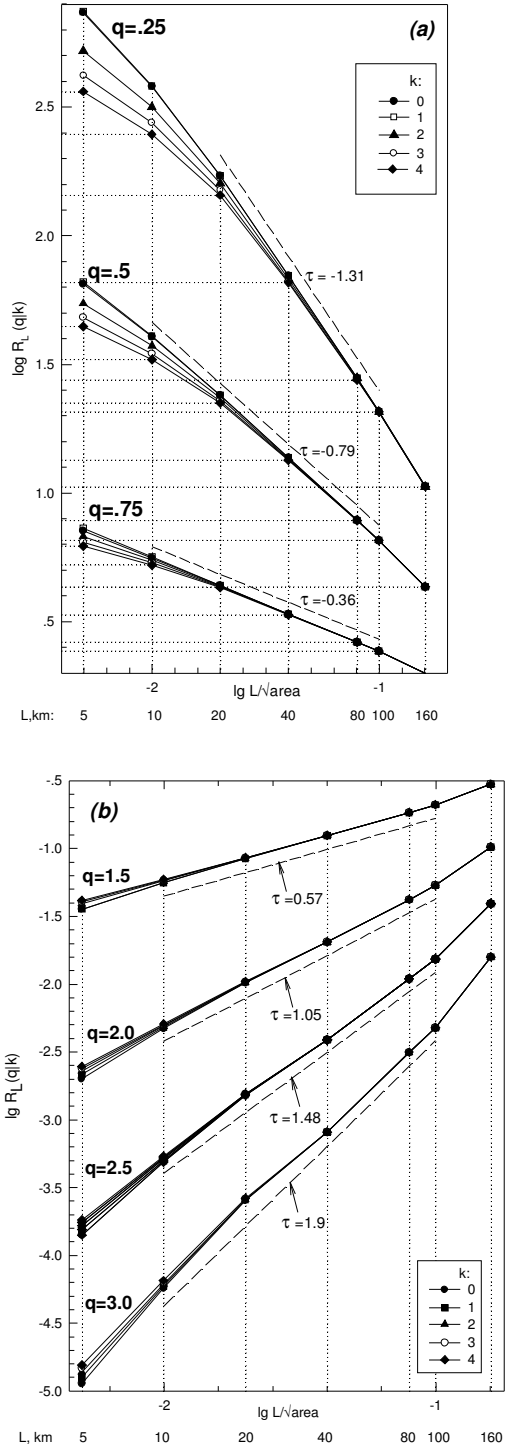

Figure 4. Data for estimating $\tau(q)$:
(a) $q = 0.25, 0.5, 0.75$; (b) $q = 1.5, 2.0, 2.5, 3.0$.
The dashed lines mean the same things as in Fig. 3. The vertical axis shows values of the modified Renyi function $R_L(q\,|\,k)$ (see Section 2.4) based on $L{\times}L$ cells with numbers of events greater than $k = 0, 1, 2, 3, 4$.

Of special interest for the present study are estimates of the derivatives $\dot\tau(q)$ at $q = 0$ and $q = 1$ (see (14)). These were found from the relations

$$n_L^{-1}\sum_{i=1}^{n_L}\lg p_i(L) = \dot\tau(0)\lg L(1+o(1)) \qquad (20)$$

$$\sum_{i=1}^{n_L} p_i(L)\lg p_i(L) = \dot\tau(1)\lg L(1+o(1)) \qquad (21)$$

where $p_i(L) = n_i(L{\times}L,T)/n(G,T)$, $n(A,T)$ is the number of events in $A$ during time $T$. Relations (20, 21) are derived from (8) by formal differentiation with respect to $q$. The method used to estimate $\dot\tau(q)$ exactly follows the estimation of $\tau(q)$. It is illustrated in Fig. 5b for $M \geq 2$ and $M \geq 3$. The effect due to the use of $n(L{\times}L,T) > k$ with $k > 0$ is not unilateral with respect to the case $k = 0$ (Fig. 5 should be compared with Fig. 3 and 4 for small $L$). This allows estimation of $\dot\tau(1)$ in the range of scales $\Delta L = (5, 100)$ km for $M \geq 2$ and $(10, 100)$ km for $M \geq 3$: $\dot\tau(1) = 1.33$ ($M \geq 2$); $1.22$ ($M \geq 3$). Somewhat unexpectedly, the evaluation of $\dot\tau(1)$ is shown more stable compared with $\tau(p)$.

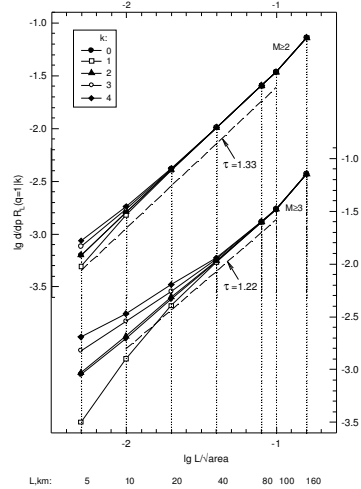

Figure 5. Data for estimating the information dimension $d_1 = \dot\tau(1)$ from $M \geq 2$ (a) and $M \geq 3$ (b) events.
The dashed lines mean the same things as in Fig. 3. The vertical axis shows the derivatives $d/dq\,R_L(q=1\,|\,k)$ for $k = 0, 1, 2, 3, 4$ where $R_L$ is the modified Renyi function (see also the left-hand side of (21)).

**Variation of the estimates.** The following quantities are of greatest interest for subsequent analysis: the box dimension $d_0 = -\tau(0)$, the correlation dimension $d_2 = \tau(2)$, $\dot\tau(0)$, and the information dimension $d_1 = \dot\tau(1)$. The respective estimates can be affected by the choice of $(L_*, L^*)$, grid location, the boundary of $G$, epicenter location uncertainty. The following options were considered for $(L_*, L^*)$: $(10, 100)$ km, $(20, 100)$ km, and $(20, 80)$ km. The grid location is specified by the center (see Fig. 1) and by the direction of the principal axis. The center was moved within $\pm 7$ km, because $L_* = 10$ km; the axis direction was varied within the limits $\pm 10°$. Along with the above region $G$ we also used an alternative one $G_1$ whose boundaries were parallel to the grid axis (see Fig. 1).

The ANSS catalog contains some poorly located events, as indicated by the number of stations used in the location procedure for an event, $n_{st}$. The estimates were varied by using two options: the events with $n_{st} \geq 1$ (the main option) and those with $n_{st} \geq 7$.

The following table sums up the variation of the estimates for the above dimensions:

| $\dot\tau(0)$ | $d_0$ | $d_1$ | $d_2$ | |
|---|---|---|---|---|
| $2 \pm 0.1$ | $1.8 \pm 0.1$ | $1.35 \pm 0.05$ | $1.1 \pm 0.05$ | (22) |

Table (22) corroborates that the dimensions $d_0$ and $d_2$ that one uses to scale $\lambda(L{\times}L)$ are significantly different.

***The distributions of*** $\xi_L$. Starting from the multifractality concept, we have arrived at the conclusion that the suitable parameter $c$ for scaling the distribution of $\lambda(L{\times}L)$ given that $\lambda(L{\times}L) > 0$ may be $c = \dot{\tau}(0) \cong 2$ and $c = d_1 \cong 1.35$. The former value is for the situation where a nonempty $L{\times}L$ cell is used with a constant weight $1/n_L$, while the latter is relevant to a weight proportional to the seismicity rate in the cell in question. Parametrically speaking, the former case corresponds to the value $p = 0$ and the latter to $p = 1$. When choosing theoretical estimates of $c$ we expect the lowest scatter in the distributions of $\xi_L$ (see (13)) in that range of scales where the measure $\lambda(dg \mid m = 2)$ behaves in a multifractal manner, i.e., when $\Delta L = (10, 100)$ km.

The distribution functions for $\lg \xi_L$ are shown in Fig. 6a (the case $p = 0$) and 6b (the case $p = 1$) for different $c$ from the list (22). The list has been supplemented with the value $c = 1.6$ corresponding to the estimate of $d_0$ by Corral (2003). The curves in Fig. 6 can also be treated as plots of the distributions of $\xi_L$ with a horizontal log axis. The choice of a log scale for $\xi_L$ is quite natural owing to two reasons: it is consistent with the meaning of the asymptotic form (6) and with large (up to five orders of magnitude) range of $\xi_L$ values (see Fig. 6).

A comparison between distributions gives rise to the issue of the appropriate metric. The Levy metric (Feller, 1966, Ch.8 §10) is quite sufficient for the case under consideration; the metric is actively employed in probability theory when examining the convergence of distributions. This metric concentrates on divergences between distributions in the region where the bulk of the distribution lies, being less sensitive to the behavior of tails. In our case small values of $\xi_L$ are related to small numbers of $n(L{\times}L,T)$, hence are very inaccurate. On the other hand, large values of $\xi_L$ involve small relative errors. However, the frequencies of occurrence for very large $\xi_L$ are low. They are supplied by aftershock sequences with high concentrations in $L{\times}L$ cells. Such sequences are few for a period of $T = 20$ years, hence the estimation of probabilities of large values of $\xi_L$ may be extremely unstable. The Levy metric can be informally defined as follows.

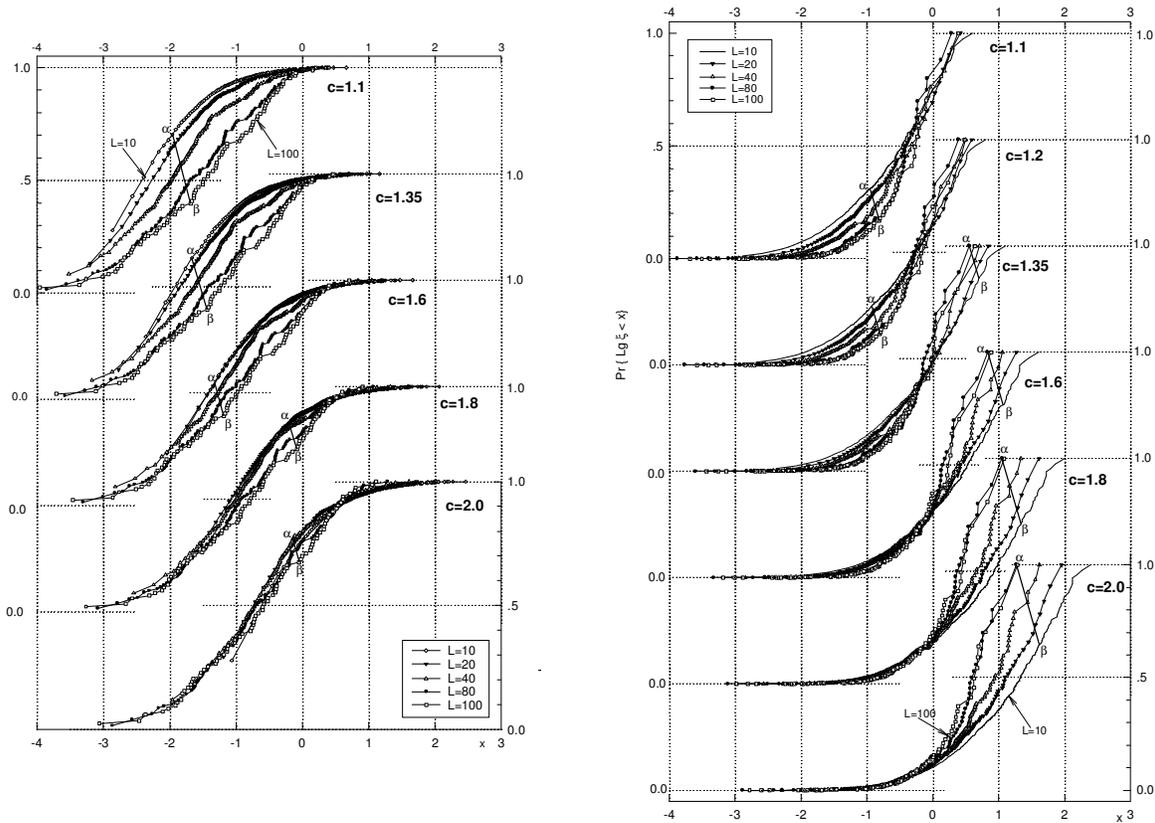

**Figure 6.** Distribution functions for $\lg \xi_L$ corresponding to the scales $L = 10, 20, 40, 80$ and $100$ km and to the scaling index $c = 1.1, 1.2, 1.35, 1.6, 1.8$ and $2.0$. The segment $(\alpha, \beta)$ has the slope $(-1)$, its length provides information on the scatter of the distributions of $\lg \xi_L$ at a fixed $c$.
Left (a): The case $p = 0$: each $L{\times}L$ cell enters in the distribution with the same weight;
Write (b):: The case $p = 1$: each $L{\times}L$ cells enters in the distribution with a weight proportional to the seismicity rate in the cell.

Consider the plots of the distribution functions $F_1(x)$ and $F_2(x)$. In the case under consideration, they correspond with the distributions of $\lg \xi_L$ for different $L$. Let us connect the plots with the help of any manner of straight segments having a common direction $(-1)$. The upper bound to the lengths of these segments is assumed to be the Levy distance between $F_1$ and $F_2$. We are considering the family of distributions of $\lg \xi_L$ with $L = 10, 20, 40, 80,$ and $100$.

The largest of the pairs of distances is taken to represent the scatter $\delta_c$ for the distributions of $\lg \xi_L$ that are related to the exponent $c$ (see segments ($\alpha, \beta$) in Fig. 6).

In Fig. 6a (the case $p = 0$) the quantity $\delta_c$ is monotone decreasing with $c$ increasing, reaching the minimum at $c = 2$, which is identical with the predicted value $c = \dot{\tau}(0)$. The case $p = 1$ involves a fine point (Fig. 6b), since the expected value $c = \dot{\tau}(1) = 1.3 - 1.4$ lies within the interval (1, 2). As $c$ increases, the greatest discrepancy between the distributions is gradually getting from small $\xi_L$ towards larger $\xi_L$. The position of equilibrium is reached in the interval $1.1 \leq c \leq 1.35$, formally at the point $c = 1.2$ where the minimum of $\delta_c$ is poorly pronounced. Even though the latter estimate of $c$ is rather diffuse, it can still be asserted that the two estimates based on different weights of $L \times L$ cells are significantly different. We have to remind that they ought to be identical for monofractal seismicity.

Figure 6(a, b) thus fairly well corroborates our theoretical analysis, and for this reason provides independent evidence in favor of a multifractal behavior of $\lambda(dg \mid m = 2)$ in the range of scales $\Delta L = (10-100)$ km. From Fig. 6a it also follows that $c = 2$ is a better candidate for the suitable value of $c$ needed to scale distributions of $\lambda(L \times L)$ than is $c = 1.6$ proposed by Corral (2003).

### 4. Conclusion.

We started by discussing the question of the best scaling to be applied to the rate of $M \geq m$ events in an $L \times L$ cell, namely, the question of a suitable exponent $c$ in the relation $\lambda(L \times L) \propto L^c$. We assumed the hypothesis of multifractality for the measure of rate $\lambda(dg \mid m)$ to show that the problem has no unique solution and requires that the ultimate goal we are pursuing should be made more specific. For example, the averaged value of $\lambda(L \times L)$ over all nonempty cells is scaled by using $c = d_0$ where $d_0$ is the box dimension of the support of $\lambda(dg \mid m)$, while the average value of $\lambda(L \times L)$ weighted proportionally to seismicity rates in the $L \times L$ cells requires $c = d_2$ where $d_2$ is the correlation dimension. If we want the distributions of the normalized $\lambda(L \times L)/L^c$ to be close to one another for different $L$ (they may obey the unified law as the ideal case), then $c = \dot{\tau}(0)$ (see Section 2.3). That same distribution can be constructed by taking the weight of the $L \times L$ cell to be proportional to the rate in $L \times L$, i.e., in the same way as we did when finding the alternative mean of the $\lambda(L \times L)$. Then $c = d_1$ where $d_1$ is the information dimension.

The multifractality hypothesis is a controversial subject. The discussion focuses on the reliability of estimates of the scaling indices $\tau(q)$ for the measure $\lambda(dg \mid m)$. For this reason we have paid special attention to the method to be used for estimating $\tau(q)$ (see section 2.4). We used California seismicity (the ANSS catalog, 2004) to show that $\lambda(dg \mid m = 2)$ demonstrates multifractal behavior in the range of scales $\Delta L = (10-100)$ km. To be more exact, $\tau(q)$ admits of stable estimation for $0 \leq q \leq 3$ in the above range of $\Delta L$, all the dimensions listed above being different: $\dot{\tau}(0) \cong 2$, $d_0 \cong 1.8$, $d_1 \cong 1.35$ and $d_2 \cong 1.1$. Independent analysis of the distributions of $\lambda(L \times L)/L^c$ (with equal and unequal weights) provides a fair corroboration of the theoretically predicted value of $c$ at which the distribution is weakly dependent on $L$. The prediction is exact for the case of equal weights ($c = \dot{\tau}(0) \cong 2$) and is approximate otherwise, namely, $1.1 \leq c \leq 1.35$ with the theoretical value $c = \dot{\tau}(1) \cong 1.3 - 1.4$. That result makes the multifractal hypothesis more plausible in the range of scales $\Delta L = (10-100)$ km.

For opponents of multifractality one can express oneself in a different manner: the result shows that the multifractal formalism is effective in solving the problem of the spatial scaling of seismicity rate. An important place is occupied by the box counting approach which was used to estimate fractal dimensions. The approach well matches the problem we are considering, because both of these cases are concerned with values of $\lambda(dg \mid m)$ in squares of size $L$ belonging to a rectangular grid. Quite independent of any interpretation to be put on the final result, one can draw the following practical inference: in situations where seismicity is scaled over space, the exponent $c$ must be treated as a parameter. At present the scaling is used in the analysis of unified laws (Bak et al., 2002; Corral, 2003), in certain prediction techniques (Baiesi, 2004), and for aftershock identification (Baiesi and Paczuski, 2004a,b).

We did not try to address the question of how the parameter $c$ or, in particular, the box dimension $d_0$, depend on the cutoff magnitude $m$. A rigorous solution encounters great difficulties. As $m$ increases, the straightforward analysis of scaling indices becomes difficult for statistical reasons (see the example in Fig. 3). When large events are concerned, source dimension should be taken into account. For this reason, similarity considerations will call for greater sampling area, and this will lead to problems with catalogs. The theory developed in Gorshkov et al. (2003) as to the occurrence of large earthquakes on high rank lineaments and their intersections provides an indirect indication that the dimension $d_0$ must decrease with increasing magnitude. On an earthwide scale great earthquakes occur at plate boundaries, and the plate dimension in the same scale is naturally associated with 1, when one deals with intersections of plates with the Earth's surface. On the whole the above hypothesis calls for serious statistical testing.

**Acknowledgements.** The final version of this paper significantly differs from the original one thanks to constructive criticisms due to the anonymous reviewers and our colleagues V. Kossobokov and M. Shnirman. This work was supported by the the Russian Foundation for Basic Research (grant 05-05-64384-a) and by the EU 6-th Framework STReP project "E2-C2 extreme events: causes and consequences".

### References

Abe, S. and Suzuki, N., 2004. Scale–invariant statistics of the degrees of separation in directed earthquake network, *ArXiv: cond–mat / 0402226*.

ANSS composite earthquake catalog, 2004. *quake.geo.berkeley.edu/anss*.

Baiesi, M., 2004. Scaling and precursor motifs in earthquake networks, *ArXiv: cond–mat / 0406198*.

Baiesi, M. and Paczuski, M., 2004a. Scale free networks of earthquake and aftershocks, *Phys. Rev. E., 69*, 066106.

Baiesi, M. and Paczuski, M., 2004b. Complex networks of earthquake and aftershocks, *ArXiv: physics / 0408018*.

Bak, P., Christensen, K., Danon, L. and Scanlon, T., 2002. Unified scaling law for earthquakes, *Phys. Rev. Lett., 88*, 178501.

Corral, A., 2003. Local distributions and rate fluctuations in a unified scaling law for earthquakes, *Phys. Rev. E., 68*, 035102 (R).

Corral, A., 2004. Universal local versus unified global scal-ing laws in the statistics of seismicity, *ArXiv: cond–mat /* 0402555.

Eneva, M., 1996. Effect of limited data sets in evaluating the scaling properties of spatially distributed data: an example from mining–induced seismic activity, *Geophys. J. Int., 124*, 773–786.

Feder, J., 1988. *Fractals*, Plenum Press, New York, 283 pp.

Feller, W., 1966. *An Introduction to Probability Theory and Its Applications*. Vol.2. John Wiley &Sons, Inc. New York.

Frisch, U., 1996. *Turbulence: the legacy of A.N. Kolmogorov*, Cambridge :Cambridge University press., 296 pp.

Geilikman, M.B., Golubeva, T.V. and Pisarenko V.F., 1990. Multifractal patterns of seismicity, *Earth Planet. Sci. Lett., 99,* N 1/2, 127–132.

Godano, C., Tosi, P., De Rubeis, V. and Augliera, P., 1999. Scaling properties of the spatio–temporal distribution of earthquakes: a multifractal approach applied to a California catalogue, *Geophys. J. Int., 138*, 99–108.

Goltz, C., 1997. *Fractal and chaotic properties of earthquakes*, Springer, Berlin.

Gonzato, G.,Mulargia, F. and Marzocchi, W.,1998. Practical application of fractal analysis: problems and solutions, *Geophys. J. Int., 132*, 275–282.

Gorski, F.Z., 2001. Pseudofractals and the box counting algorithm, *J. Phys. A.: Math.Gen., 34*, 7933–7940.

Gorshkov A., Kossobokov V. and Soloviev A., 2003. Recognition of earthquake–prone areas. *In* Keilis–Borok V.I. and Soloviev A.A. (eds.) *Nonlinear dynamics of the lithosphere and earthquake prediction.* Springer. 239–310.

Grassberger, P. and Procaccia, I., 1983. Measuring the strangeness of strange attractors *Phisica D, 9,* 189–208.

Hirabayashi, I., Ito, K. and Yoshii, T., 1992. Multifractal analysis of earthquakes, *Pure appl. geophys., 138*, 591–610.

Hirata, T. and Imoto, M., 1991. Multifractal analysis of spatial distribution of microearthquakes in the Kanto region, *Geophys. J. Int., 107*, 155–162.

Kagan, Y., 1991. Fractal dimension of brittle fracture *J. Nonlinear Sci., 1,* 1–16.

Keilis–Borok, V.I., Kossobokov, V.G. and Mazhkenov,S.A., 1989. On similarity in spatial seismicity distribution, *Computational Seismology, 22,* 28–40, Moscow, Nauka.

Kossobokov, V.G. and Nekrasova, A.K., 2004. A general similarity law for earthquakes: a worldwide map of the parameters, *Computational Seismology, 35,* 160–175, Moscow, GEOS.

Mandelbrot, B.B., 1989. Multifractal measures for the geophysicist, *Pure appl. geophys., 131*, 5–42.

Molchan, G.M., 1996. Scaling exponents and multifractal dimensions for independent random cascades, *Commun. Math. Phys., 179*, 681–702.

Molchan, G.M., 2005. Interevent time distribution of seismicity: a theoretical approach, *Pure appl. geophys., 162*, 16 p.

Molchan, G.M. and Turcotte, D., 2002. A stochastic model of sedimentation: probabilities and multifractality, *Euro. Jnl. of Applied Mathematics, 13*, 371–383.

Nerenberg, M.A. and Essex, C., 1990. Correlation dimension and systematic geometric effects, *Phys. Rev. A, A42*, 7065–7074.

Pesin, Y. and Weiss, H., 1997. A multifractal analysis of equilibrium measures for conformal expanding maps and Moran–like geometric constructions, *J. Stat. Phys., 86*, Nos.1/2, 233–275.

Pisarenko V.F. and Golubeva, T.V., 1996. Application of stable laws in seismicity modeling, *Computational seismology and Geodynamics, 4*, 127–137.

Tosi, P., De Rubeis, V., Loreto, V. and Pietronero, L., 2004. Influence length and space–time correlation between earthquakes, *ArXiv: physics /* 0409033.